\documentclass[conference, 10pt]{IEEEtran}
\usepackage{cite}
\usepackage{url}
\usepackage{graphicx}
\usepackage{color}
\usepackage{placeins}
\usepackage{float}
\usepackage{tabularx,colortbl}
\usepackage{ifthen}
\usepackage{subcaption}
\usepackage{psfrag}

\usepackage{epsfig,wrapfig}
\usepackage{epstopdf} % for pdflatex
\usepackage{graphicx}
\usepackage{fancyhdr}
\usepackage{newtxmath}
\usepackage{caption}
\usepackage{subcaption}
\usepackage{psfrag}
\makeatletter

\newcounter{author}
\renewcommand{\author}[2][]{
	\stepcounter{author}
	\@namedef{author@\theauthor}{#2}
	\@namedef{authorlabel@\theauthor}{#1}
}

\newcounter{address}
\newcommand{\address}[2][]{
	\stepcounter{address}
	\@namedef{address@\theaddress}{#2}
	\@namedef{addresslabel@\theaddress}{#1}
}

\newcommand{\alsep}{and}

\def\newmaketitle{\par%
	\begingroup%
	\normalfont%
	\def\thefootnote{}%  the \thanks{} mark type is empty
	\def\footnotemark{}% and kill space from \thanks within author
	\let\@makefnmark\relax% V1.7, must *really* kill footnotemark to remove all \textsuperscript spacing as well.
	\footnotesize%       equal spacing between thanks lines
	\footnotesep 0.7\baselineskip%see global setting of \footnotesep for more info
	\normalsize%
	\twocolumn[\thenewmaketitle\@IEEEaftertitletext]%
	\if@IEEEusingpubid
	\enlargethispage{-\@IEEEpubidpullup}%
	\fi
	\endgroup
	\setcounter{footnote}{0}\let\maketitle\relax\let\@maketitle\relax
	\gdef\@thanks{}%
	\let\thanks\relax}

\def\thenewmaketitle{
	\newpage
	\begin{center}%
		\vskip0.2em{\Huge\@IEEEcompsoconly{\sffamily}\@IEEEcompsocconfonly{\normalfont\normalsize\vskip 2\@IEEEnormalsizeunitybaselineskip
				\bfseries\large}\@title\par}\vskip1.0em\par%
		\vspace{1ex}
		\newcounter{c@author}
		\newcounter{c@tmp}
		\ifthenelse{\value{author}=2}{%
			\newcommand{\liand}{ and }}{%
			\newcommand{\liand}{, and }}
		\ifthenelse{\value{address}<2}{%
			\@nameuse{author@1}%
			\stepcounter{c@author}%
			\whiledo{\value{c@author}<\value{author}}{%
				\setcounter{c@tmp}{\value{author}}%
				\addtocounter{c@tmp}{-\value{c@author}}%
				\ifthenelse{\value{c@tmp}=1}{%
					\renewcommand{\alsep}{\liand}}{\renewcommand{\alsep}{, }}%
				\stepcounter{c@author}\alsep \@nameuse{author@\thec@author}}\\%
		}
		{%Add address references after the author's name
			\@nameuse{author@1}${}^{(\ref{\@nameuse{authorlabel@1}})}$%
			\stepcounter{c@author}%
			\whiledo{\value{c@author}<\value{author}}{%
				\setcounter{c@tmp}{\value{author}}%
				\addtocounter{c@tmp}{-\value{c@author}}%
				\ifthenelse{\value{c@tmp}=1}{%
					\renewcommand{\alsep}{\liand}}{\renewcommand{\alsep}{, }}%
				\stepcounter{c@author}\alsep \@nameuse{author@\thec@author}%
				${}^{(\ref{\@nameuse{authorlabel@\thec@author}})}$%
			}
		}
		\vspace{0.2ex}
		
		\ifthenelse{\value{address}>0}{%
			\ifthenelse{\value{address}=1}{
				{\@nameuse{address@1}}
			}
			{%Output the addresses as an enumerated list
				\newcounter{c@address}
				
				\begin{center}
					\whiledo{\value{c@address}<\value{address}}
					{
						\refstepcounter{c@address}
						${}^{(\thec@address)}$\,%
						\label{\@nameuse{addresslabel@\thec@address}}%
						\@nameuse{address@\thec@address}\\ %
					}
				\end{center}
			} % end of the address creation ifthenelse block
		}
		{
			\relax
		}
	\end{center}
}

\makeatother

\title{Dispersion Characteristics of Accelerated Spacetime-Modulated Media}

\author[org1]{Amir Bahrami}
\author[org1]{Christophe Caloz}
\address[org1]{Department of Electrical Engineering, KU Leuven, Leuven, Belgium}

\begin{document}
	
	\newmaketitle
	
	\begin{abstract}
		This paper opens up the field of nonuniform-velocity SpaceTime-Modulated (STM) metamaterials, with the canonical example of an STM metamaterial of constant proper acceleration or, equivalently, hyperpolic acceleration. Combining tools of General Relativity and Classical Electrodynamics, it derives the dispersion relation of this exotic medium and reports its fundamental physics, whose most striking feature is the bending of light in the direction opposite to the direction of the modulation.
	\end{abstract}
	
	\section{Introduction}
	
	Spacetime-modulated (STM) media are dynamic structures whose constitutive parameters are modulated in both space and time~\cite{caloz2019spacetime1,caloz2019spacetime2}. In contrast to moving media, they do not involve any net transfer of matter (atoms and molecules), but a wave perturbation that travels along the background material. The modulation is often periodic, and the medium may operate in the Bragg (bandgap) regime or in the subwavelength (metamaterial) regime~\cite{huidobro2021homogenization}. STM media have many applications, including as isolation~\cite{yu2009complete}, amplification~\cite{cassedy1963dispersion,cassedy1967dispersion} and equivalent Fresnel-Fizeau light deflection~\cite{huidobro2019fresnel}.
	
	The vast majority of the STM media studied to date have been restricted to uniform-velocity modulation~(e.g., \cite{deck2019uniform}). Nonuniform-velocity -- or accelerated -- STMs, which may be considered as electromagnetic counterparts of gravitational physics~\cite{misner1974gravitation}, have been essentially unexplored. We report here a initial research step in this area by presenting the dispersion characteristics of an accelerated STM metamaterial with constant proper acceleration or, equivalently, with hyperbolic acceleration.
	
	\section{Hyperbolic Accelerated STM Medium}
	
\begin{figure}[h!]
		\centering
		\vskip3em
		\psfrag{ctt}{$ct'$}
		\psfrag{ct}{$ct$}
		\psfrag{x}{$x$}
		\psfrag{z}{$z$}
		\psfrag{zz}{$z, z'$}
		\psfrag{n1}[c][c]{$\epsilon_1$}
		\psfrag{n2}[c][c]{$\epsilon_2$}
		\psfrag{n3}[c][c]{$\mu_1$}
		\psfrag{n4}[c][c]{$\mu_2$}
		\vspace{-1.2cm} \hspace*{-0.0cm}\subfloat[] {\includegraphics[width=0.23\textwidth,height=0.16\textwidth]{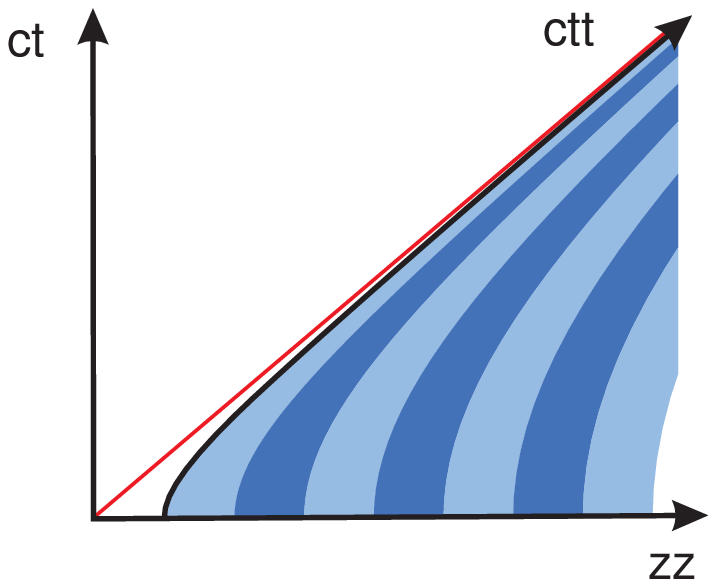}\label{fig:Rindler_Minkowski}}
		\hfill
		\subfloat[]{\includegraphics[width=0.23\textwidth,height=0.16\textwidth]{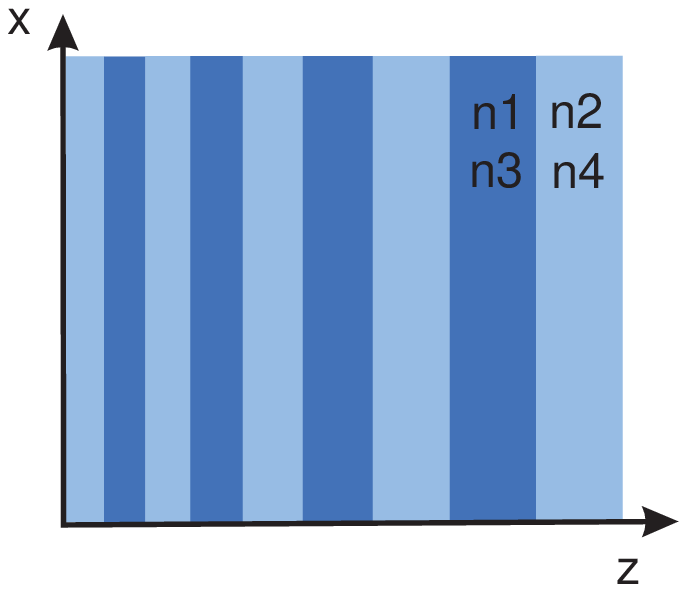}\label{fig:Spatial_plane}}
		\caption{STM metamaterial with hyperbolic (constant proper) acceleration. (a)~$z$-$ct$ plane perspective.
			(b)~$z$-$x$ plane perspective.}
		\vspace{-1cm}
		\label{fig:Rindler}
	\end{figure}

	Figure~\ref{fig:Rindler} depicts the hyperbolic accelerated STM metamaterial. The modulation is composed of layers with alternating permittivity and permeability $(\epsilon_1,\mu_1)$ and $(\epsilon_2,\mu_2)$ whose spatial/temporal widths are sufficiently smaller than the wavelength/period of the (monochromatic plane) wave to be processed to warrant homogenization, which corresponds to the metamaterial regime. 
	
	Figure~\ref{fig:Rindler_Minkowski} shows the spacetime diagram of the medium, where the constant proper acceleration\footnote{Constant proper acceleration is a type of acceleration where the observer in the comoving frame (instantaneous rest frame) experiences a constant force.}, $a'=\text{const.}$, curves, assumed here to all have the same initial velocity $v_0$ (inverse of slope at $t=0$ in Fig.~\ref{fig:Rindler_Minkowski}), are seen as hyperbolas in the laboratory frame~\cite{moller1952theory}. Figure~\ref{fig:Spatial_plane} shows the double-space representation of the medium at a given time, which corresponds to a horizontal section of the spacetime graph in Fig.~\ref{fig:Rindler_Minkowski}, revealing the accelerated metamaterial is de facto nonuniformly periodic in the laboratory frame. 
	
	The problem at hand will analyzed by leveraging the basic approach of general relativity. The first step of this approach is to determine the relation between the laboratory frame (unprimed coordinates) and the comoving frame (primed coordinates) in Fig.~\ref{fig:Rindler_Minkowski}. This relation is provided by the Rindler transformation equations,
	\begin{subequations}\label{eq:Rinder_Transformations}	
		\begin{equation}
			z=\frac{c^2}{a'}\left(1+\frac{a'z'}{c^2}\right) \cosh (\xi+\xi_0)-\frac{c^2}{a'}\cosh(\xi_0), \quad
			x=x',
		\end{equation}
		\begin{equation}
			ct=\frac{c^2}{a'}\left(1+\frac{a'z'}{c^2}\right) \sinh(\xi+\xi_0)-\frac{c^2}{a'}\sinh(\xi_0), 
		\end{equation}
	\end{subequations}
	where $\xi=a't'/c$, and $\xi_0=\sinh^{-1}(\gamma_0 \beta_0)$, with $\beta_0=v_0/c$ and $\gamma_0=1/\sqrt{1-\beta_0^2}$  being the initial ($t=0$) velocity and Lorentz factor, respectively.
	
	\section{Dispersion Relation}
	In the comoving frame, the modulation is stationary, but the background material is moving towards the $-z$-direction, which induces bianisotropy\footnote{Note that this moving modulation situation is different from that of moving matter, where bianisotropy occurs in the laboratory frame~\cite{kong1990electromagnetic}, whereas it occurs here in the comoving frame.}. We have thus a bianisotropic stratified medium. Given the assumed metamaterial regime of the structure, we average the corresponding bianisotropic constitutive parameters, which gives rise to distinct permittivity, permeability and magnetoelectric coupling quantities in the $z'$- and $x'$-directions. The dispersion relation in the comoving frame  is then found from $k'_\mu k'^\mu=0$~\cite{van2012relativity} as
	\begin{equation}\label{eq:dispersion_K'}
		\frac{(k_z'+\omega' \overline{\chi'}/c)^2}{\overline{\epsilon'_x} \overline{\mu'_y}}+ \frac{k_x'^2}{\overline{\epsilon'_z} \overline{\mu'_y}}=g_{00}\left( \omega'/c \right)^2,
	\end{equation} 
	where $\overline{\epsilon'_x}=(\epsilon_1 \alpha'_1 +\epsilon_2 \alpha'_2 )/2$, $\overline{\mu'_y}=(\mu_1 \alpha'_1 +\mu_2 \alpha'_2 )/2$, $\overline{\epsilon'_z}^{-1}=(\epsilon_1^{-1}+\epsilon_2^{-1})/2$,$\overline{\chi'}=(\chi'_1+\chi'_2)/2$, $\alpha'_{1,2}=(1-\beta^2)/(1-\beta^2 \epsilon_{1,2} \mu_{1,2})$, and $\chi'_{1,2}=\beta (1-\epsilon_{1,2} \mu_{1,2})/(1-\beta^2 \epsilon_{1,2} \mu_{1,2})$, and $g_{00}=\left( 1+(a'z'/c^2)\right) ^2$.
	The relation~\eqref{eq:dispersion_K'} is then transposed to the laboratory frame using~\eqref{eq:Rinder_Transformations}, which yields
	\begin{equation}\label{eq:dispersion_K}
		\frac{\left( k_z-\frac{\psi \omega}{\sigma c}\right)^2}{\psi^2-\sigma \Omega}+ \frac{k_x^2}{(\psi^2-\sigma \Omega)\sigma \gamma^2 \epsilon_z \mu_y}=\left(\omega/c \right)^2,
	\end{equation} 
	where $\sigma=1-(\overline{\epsilon'_x} \overline{\mu'_y}-\overline{\chi'}^2)\beta^2-2\overline{\chi'}\beta$, $\Omega=\beta^2-(\overline{\epsilon'_x} \overline{\mu'_y}-\overline{\chi'}^2)-2\overline{\chi'}\beta$, and $\psi=\beta-\beta (\overline{\epsilon'_x} \overline{\mu'_y}-\overline{\chi'}^2) -\overline{\chi'}(1+\beta^2)$.
	
\begin{figure}[!h]
		\centering
		\vskip3em
		\psfrag{t}{$t$($\mu$s)}
		\psfrag{0}[c][c]{$0$}
		\psfrag{2}[c][c]{$2$}
		\psfrag{4}[c][c]{$4$}
		\psfrag{6}[c][c]{$6$}
		\psfrag{8}[c][c]{$8$}
		\psfrag{-2}[c][c]{$-2$}
		\psfrag{-4}[c][c]{$-4$}
		\psfrag{-6}[c][c]{$-6$}
		\psfrag{-8}[c][c]{$-8$}
		\psfrag{kx}{$k_z$}
		\psfrag{kz}{$k_x$}
		\psfrag{vm1}[c][c]{$v/c=0$}
		\psfrag{vm2}[c][c]{$v/c=0.12$}
		\psfrag{ss}[c][c]{stationary}
		\psfrag{uni}[c][c]{uniform}
		\vspace{0cm}\hspace*{0.0cm}\subfloat[]{\includegraphics[width=0.3\textwidth,height=0.3\textwidth]{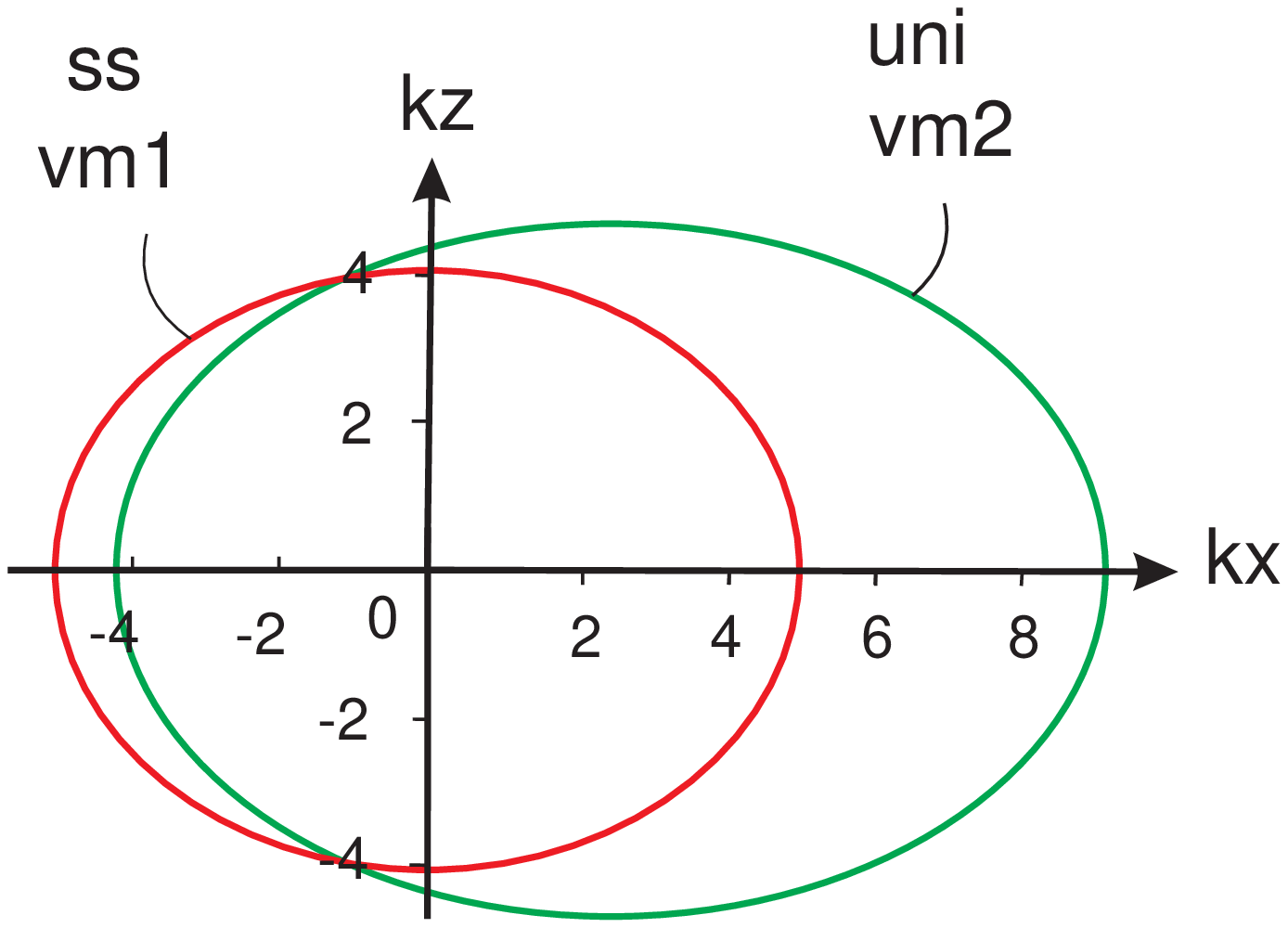} \vspace{-1.5cm} \label{fig:isofreq_Unif}} \\
		\subfloat[] {\includegraphics[width=0.4\textwidth,height=0.23\textwidth]{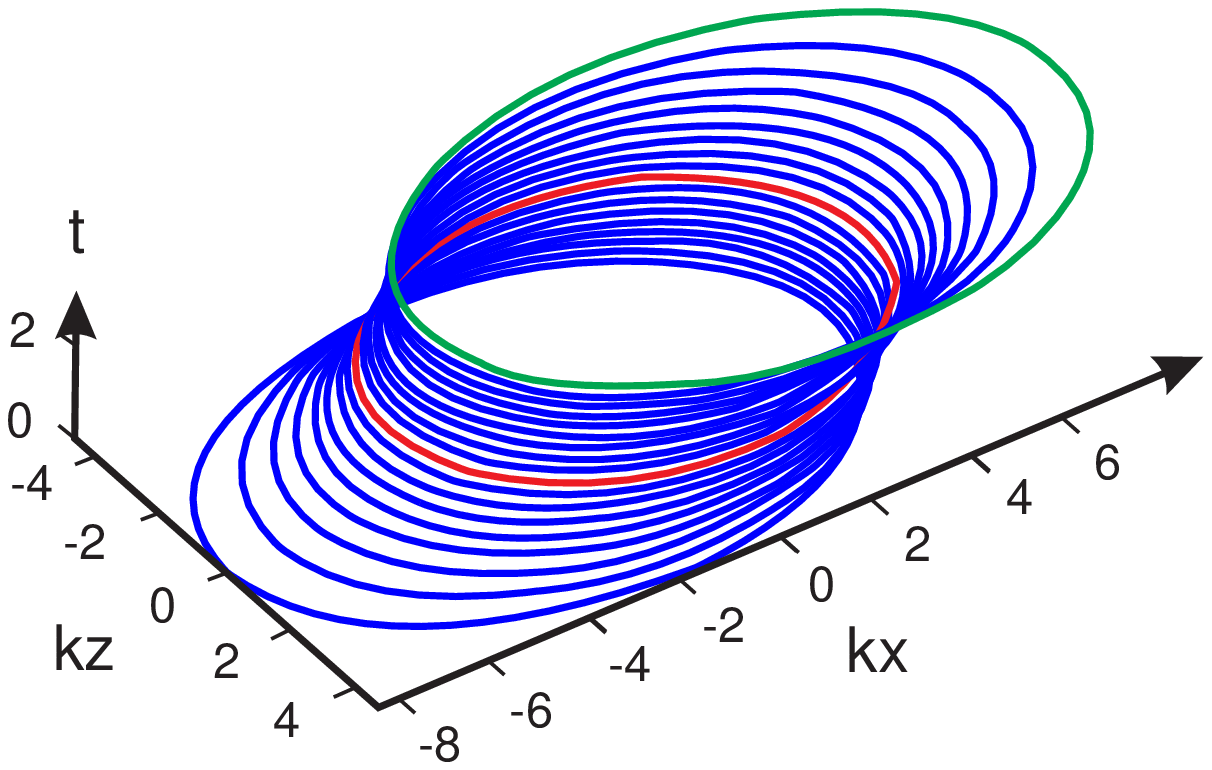}\vspace{+1cm}\label{fig:isofreq_acc}}
		\caption{Isofrequency curves for (nonmagnetic) layers of refractive indices $n_1=2$ and $n_2=8$. (a)~Uniform velocity. (b)~ Hyperbolic acceleration, with initial velocity $v_0/c=-0.12$ and acceleration $a'/c^2=0.0004$.}
		
		\label{fig:isofreq}
	\end{figure}

	Figure~\ref{fig:isofreq} plots the isofrequency curves corresponding to~\eqref{eq:dispersion_K}. Figure~\ref{fig:isofreq_Unif} compares the stationary modulation ($v=0$) and uniform-velocity modulation ($v=\text{const.}$), showing that the isofrequency curves are shifted in the $+z$-direction due to modulation\footnote{The shift would be in the $-z$-direction in the case of moving matter.}, while Fig.~\ref{fig:isofreq_acc} shows the evolution of the isofrequency curves for hyperbolic acceleration, showing their gradual shift towards $+z$ as time passes.
	
\begin{figure}
		\psfrag{s}{$\overline{S}$}
		\psfrag{0}{ }
		\psfrag{2}{ }
		\psfrag{4}{ }
		\psfrag{6}{ }
		\psfrag{8}{ }
		\psfrag{kx}{$k_x$}
		\psfrag{kz}{$k_z$}
		\psfrag{k}{$\overline{k}$}
		\psfrag{t0}{$t=0\mu \text{s}$}
		\psfrag{t3}{$t=0.5\mu \text{s}$}
		\psfrag{t2}{$t=1\mu \text{s}$}
		\begin{subfigure}{\linewidth}
			\includegraphics[width=.3\linewidth]{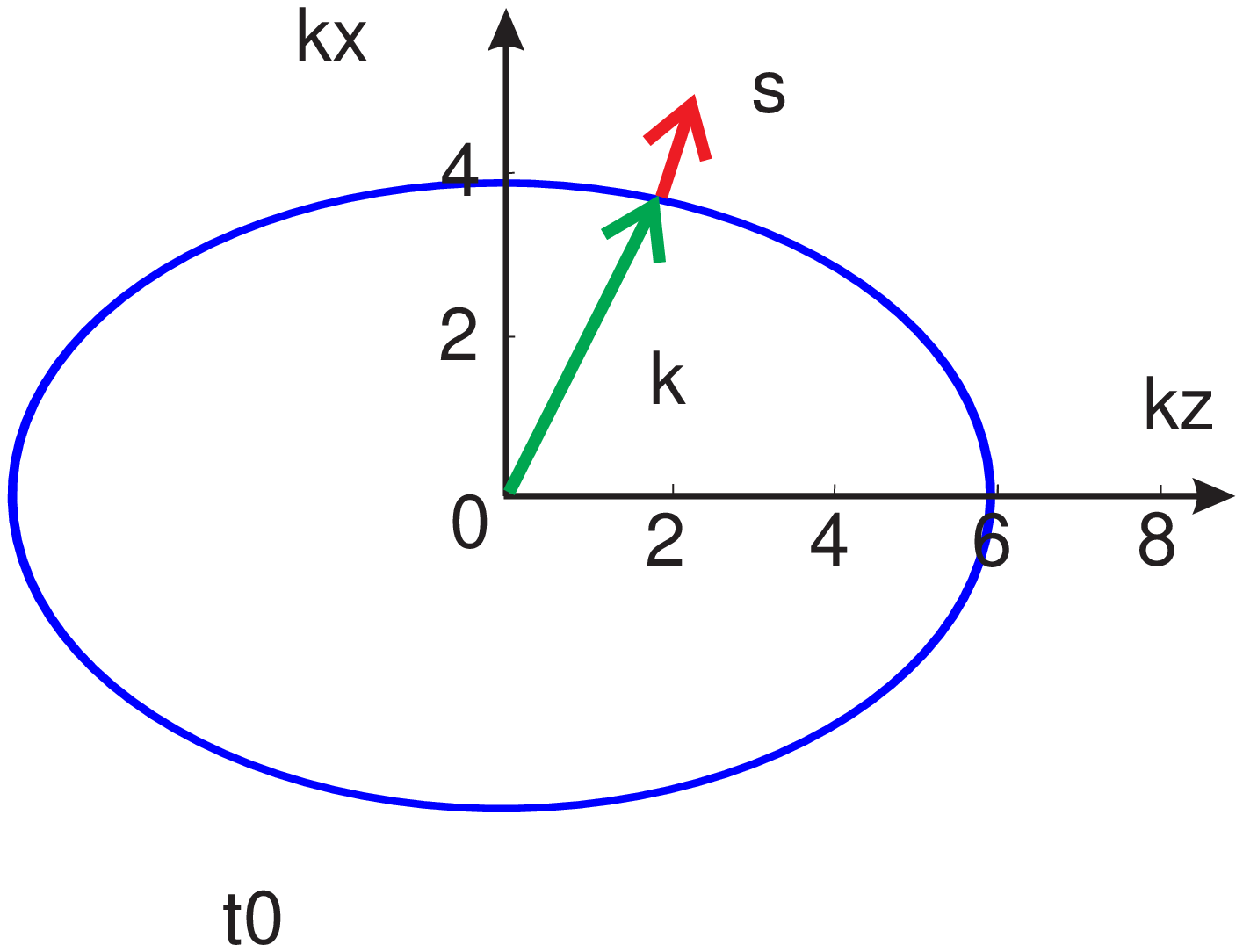}\hfill
			\includegraphics[width=.3\linewidth]{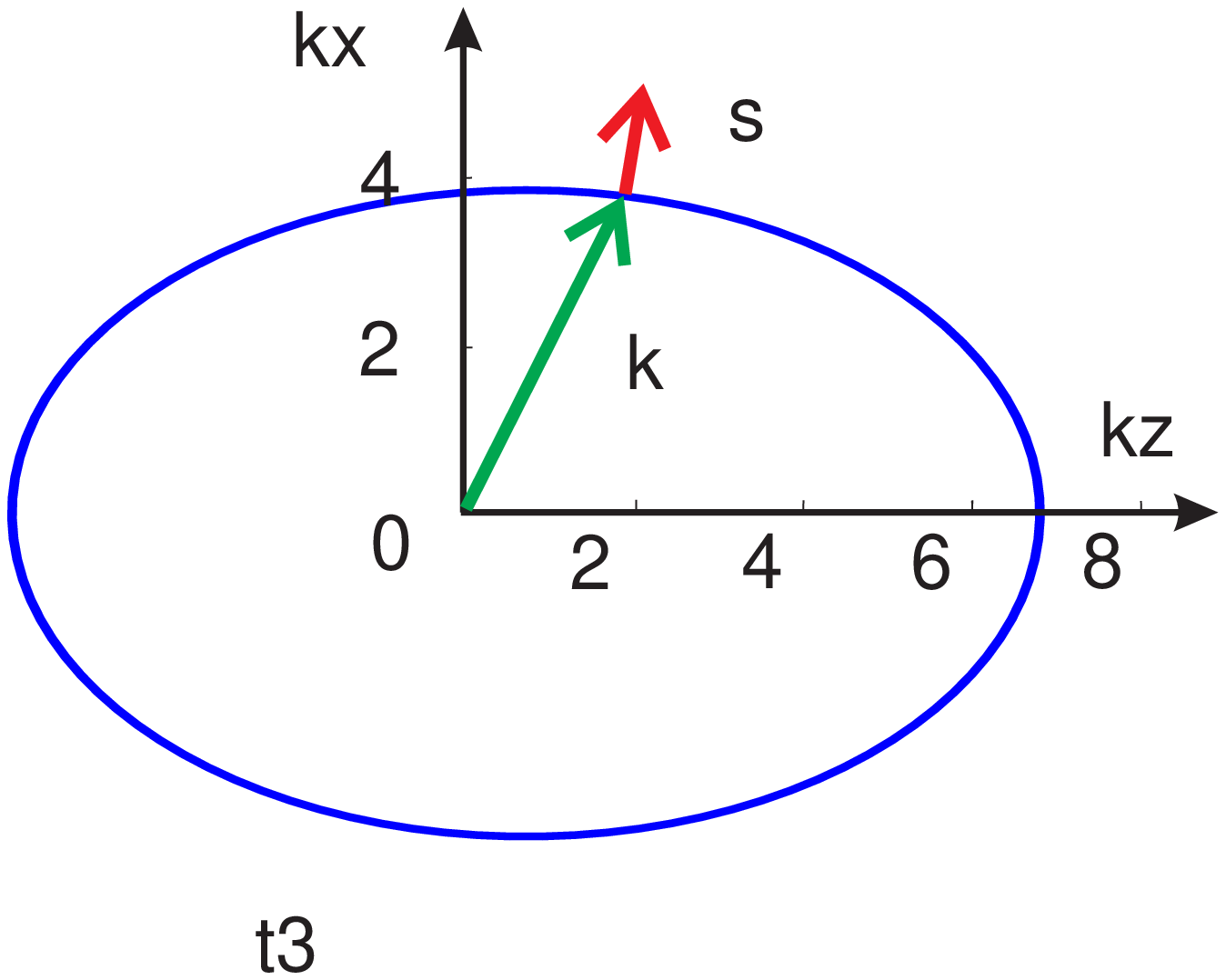}\hfill
			\includegraphics[width=.3\linewidth]{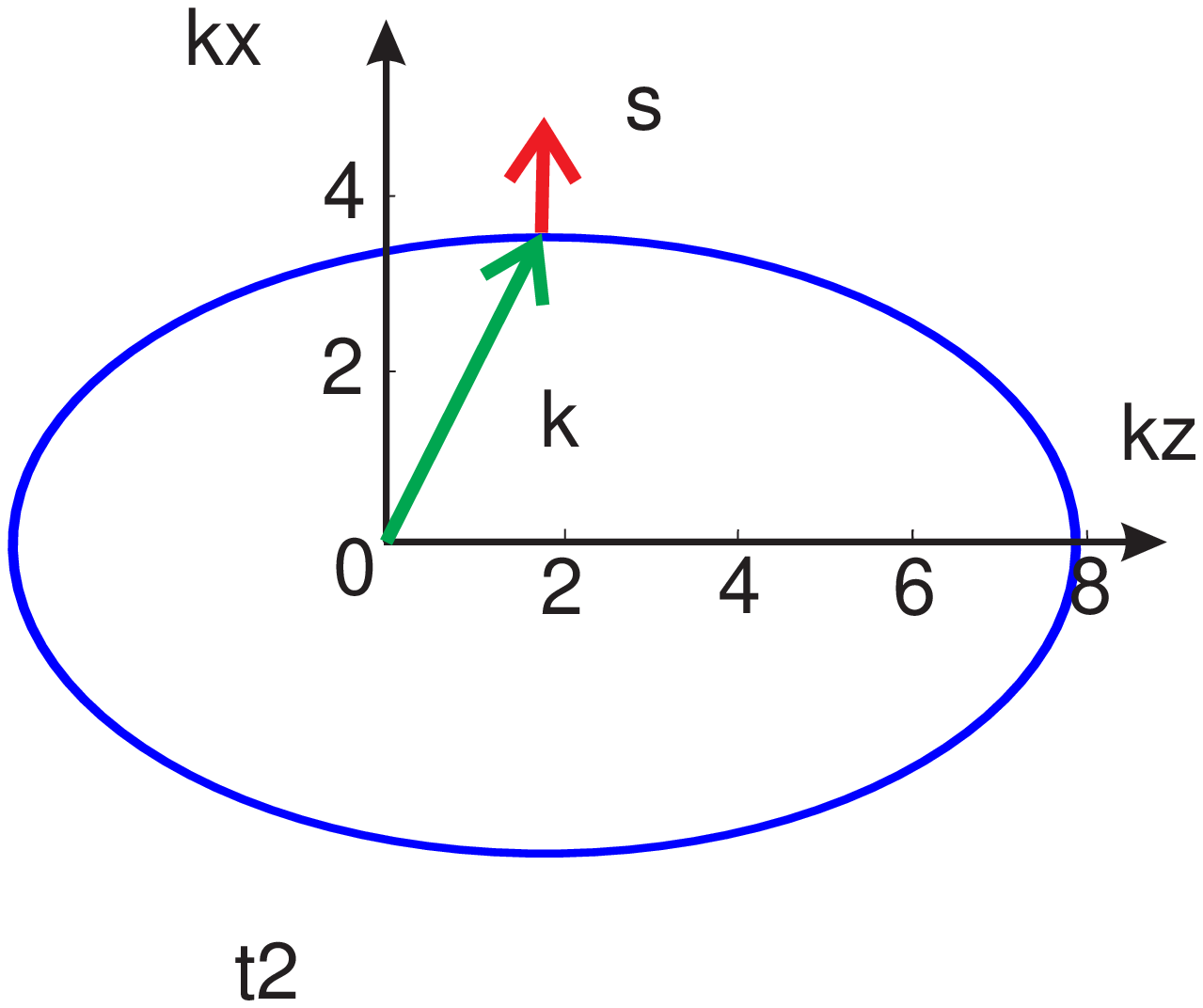}
			\caption{}\label{Fig: deflection_modulation}
		\end{subfigure}
		\begin{subfigure}{\linewidth} 
			\includegraphics[width=.3\linewidth]{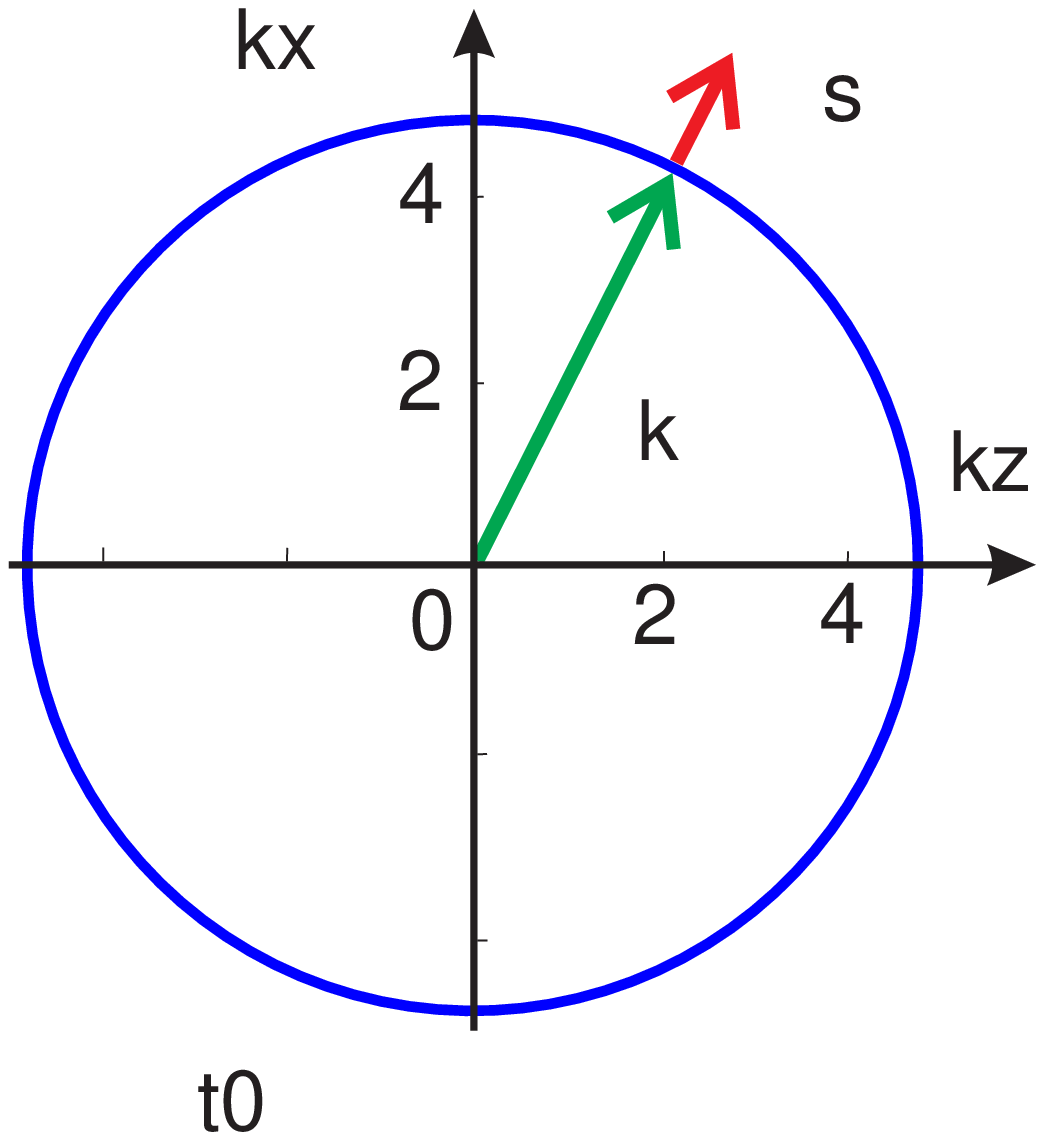}\hfill
			\includegraphics[width=.3\linewidth]{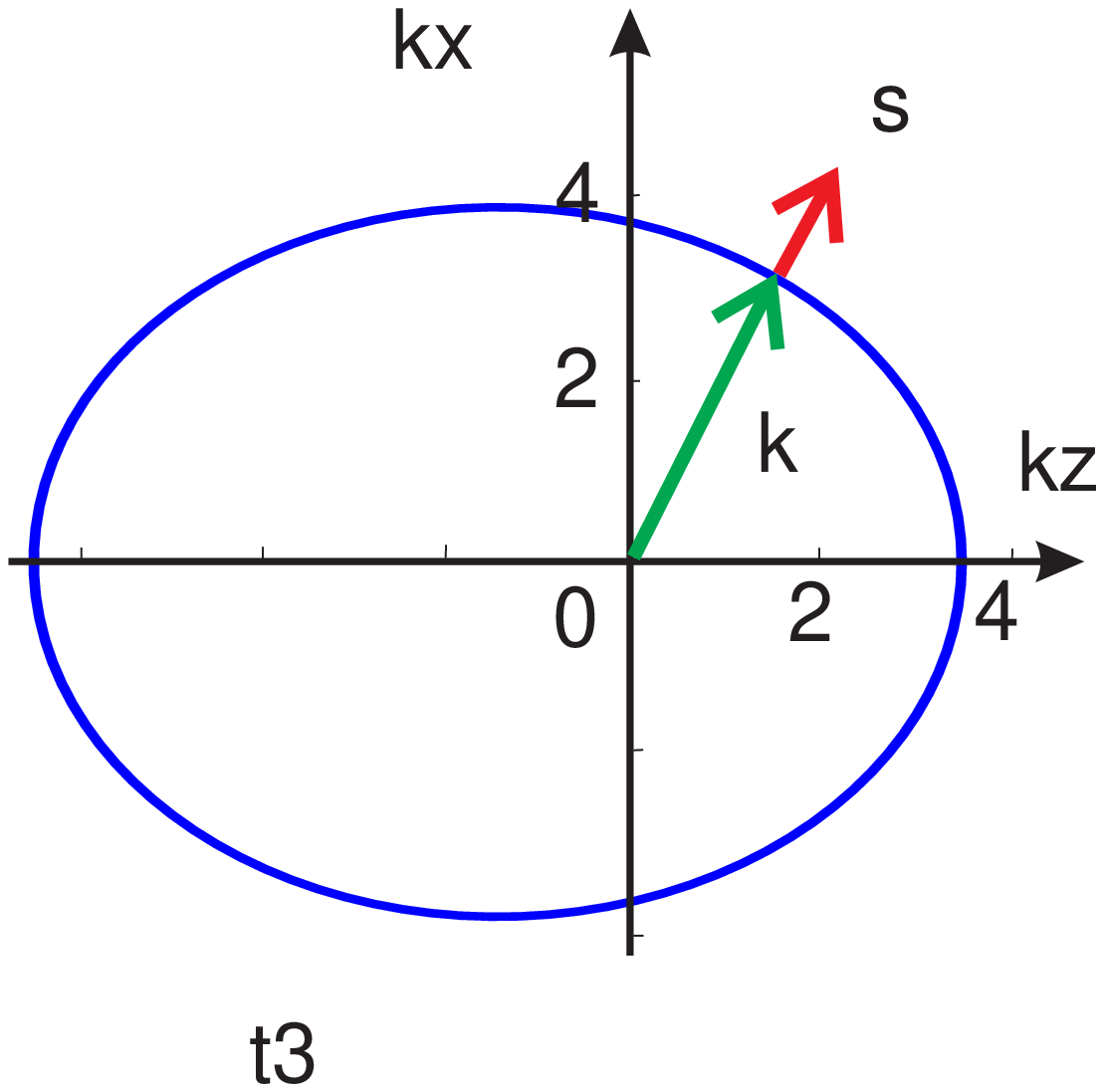}\hfill
			\includegraphics[width=.3\linewidth]{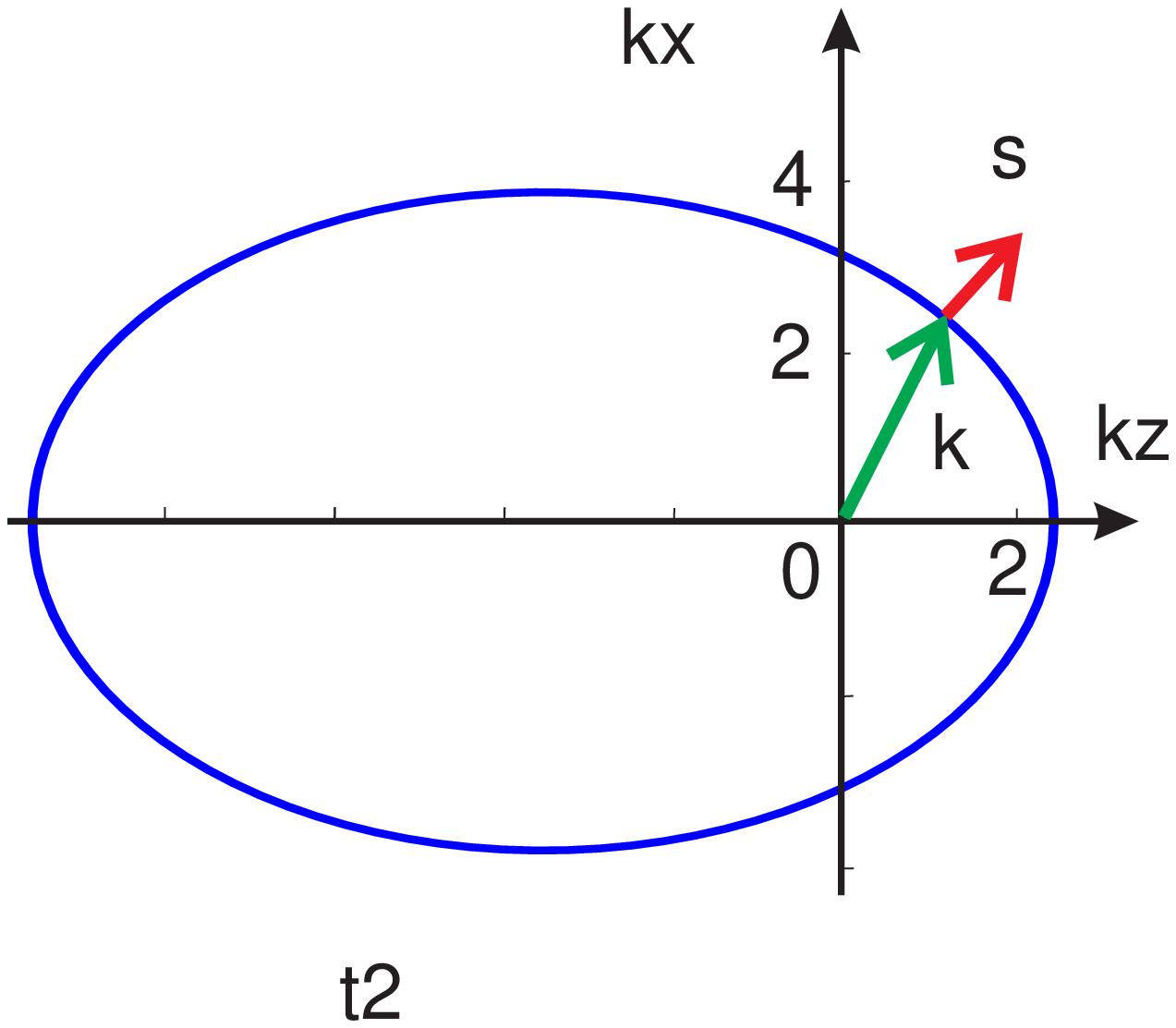}
			\caption{}\label{Fig: deflection_matter}
		\end{subfigure}
		\caption{Light bending due to acceleration. (a)~Hyperbolic accelerated STM metamaterial corresponding to Fig.~\ref{fig:isofreq_acc}. (b)~Accelerated matter (dielectric) counterpart of (a) with $n=5$.}
		\label{Fig: deflection}
	\end{figure}

	The isofrequency curves given by~\eqref{eq:dispersion_K} provide the directions of the phase and group velocities, which are parallel to $\overline{k}$ and $\overline{S}=\nabla_k \omega(\overline{k})$, respectively. Figure~\ref{Fig: deflection} compares the light bending due to acceleration for an STM metamaterial, in Fig.~\ref{Fig: deflection_modulation}, and an accelerating medium, in Fig.~\ref{Fig: deflection_matter}. The results show that light bends in the direction opposite to the perturbation motion in the former case, whereas, as well-known from Einstein's theory, it bends in the direction of matter motion in the latter case.

	% used to balance the columns on the last page adjust value as needed
	
	
	%\bibliographystyle{IEEEtran}
	
	\bibliographystyle{IEEEtran}
	\bibliography{Reference.bib}

\end{document}